\documentclass[showpacs,amsmath,amssymb,floatfix,prl,twocolumn]{revtex4}

\usepackage[dvips]{graphicx}
\usepackage{amssymb,amsfonts,amsmath}
\usepackage{graphicx,amsmath}

\begin{document}

\title{Non linear transport in drift-diffusion equations under magnetic field}
\author{A.D. Chepelianskii\\
Cavendish Laboratory, University of Cambridge, J J Thomson Avenue, Cambridge CB3 OHE, UK}


\begin{abstract}
We analyze numerically and analytically the non linear transport properties 
of a drift-diffusion equation in presence of a magnetic field and of a 
disorder potential. For a wide range of parameters this model
exhibits a plateau where the drift velocity is almost independent on the applied electric field.
This behavior has strong similarities with the zero differential resistance 
states observed experimentally in high mobility two dimensional systems. 
Performed numerical simulations are in a good global agreement
with the developed analytical theory even if the later leads to overestimated 
negative differential resistance values. 
\end{abstract}

\maketitle

The study of out of equilibrium transport in high purity two dimensional electron systems 
has lead to the discovery of several surprising effects \cite{zudov2001a,ye2001,yang2002,zudov2001b,dorozhkin2005,studenikin2005,zudov2006,mani2010} 
and stimulated an intense theoretical research \cite{ryzhii,girvin,polyakov,vavilov,platero,platero2,mirlin,monarkha}.
Probably the most striking phenomenon observed so far is the formation 
of zero-resistances states under microwave irradiation at certain ratios between 
microwave and cyclotron frequencies \cite{mani2002,zudov2003,mani2004,smet,bykov,bykov2009,Wiedmann}.
It is widely argued that in a zero resistance state the electron gas undergoes 
a transition to a state where the Hall current splits into several domains \cite{girvin,polyakov,vavilov,mirlin,monarkha}.
Zero resistance is then explained as the result of a cancellation between 
the contributions of the different domains. 
Although a microscopic theory describing the domain formation remains unavailable,
the domain concept was used to interpret many other experiments.
Thus when zero differential resistance states (ZDRS) were first reported,
it was assumed that they were also a manifestation of this domain physics \cite{bykov2007,zhang2007,kunold2009,zudov2010,gusev2011,bykov2011}. 
However recent experiments on ZRS in liquid Helium show that the 
steady state of the surface electrons in this regime 
is not a domain state but an accumulation of the electrons 
on the edges of the electron cloud \cite{denis2010,denis2011}. This and recent alternative 
theoretical proposals which do not rely on the domain picture \cite{toulouse,mikhailov,toulousecam} ,
stimulated us to see if an alternative explanation was possible for ZDRS as well. 
Thus in this letter we propose a simple transport model which produces an effect very similar to ZDRS
without appealing to a domain formation picture.

We will start by introducing the transport model employed in our investigations.
A carrier in a two dimensional electron gas moves under the influence 
of a disorder potential $U(\mathbf{r}) = \int \frac{d^2 q}{(2\pi)^2} {\hat U}(\mathbf{q}) e^{i \mathbf{q} \mathbf{r}}$,
electron-electron and electron-phonon interactions. 
According to results from classical dynamics \cite{chirikov}, an
electron will follow adiabatically all the fluctuations of the potential on a wavelength $2 \pi/q$
much larger than the Larmor radius when placed under a magnetic field. 
Thus we separate the disorder potential into two independent contribution, the long wavelength part 
which we henceforth call $U(\mathbf{r})$ for simplicity and a short wavelength contribution.
Our next assumption is that the short-wavelength disorder, electron-phonon and electron-electron 
interactions give rise to a ``microscopic'' longitudinal mobility $\mu_{xx}$ and to a ``microscopic'' 
diffusion constant $D$. Under these assumptions the simplest model to describe electron
transport is a drift-diffusion equation:
\begin{align}
\partial_t P = {\rm div} \left( {\hat \mu} [ \mathbf{{\rm grad}}\;U - \mathbf{E} ] P \right) + D \Delta P 
\label{Eq1}
\end{align}
where $P(\mathbf{r})$ is the probability distribution of the electron, 
${\hat \mu}$ is the mobility tensor with diagonal and off-diagonal components $\mu_{xx}$ and  $\pm \mu_{xy}$ respectively. 
We have also introduced the applied static electric field: $\mathbf{E}$. To specify the model completely 
we need to define the correlator $\Delta(\mathbf{r}-\mathbf{r'}) = <U(\mathbf{r}) U(\mathbf{r'})>$. 
In two dimensions a possible form 
for its Fourier transform is $\Delta(q) = \frac{2 \pi \lambda^2}{Q^2} \exp\left( - \frac{q^2}{2 Q^2} \right)$,
where $\lambda$ gives the disorder amplitude and the scale $Q^{-1}$ determines the typical wavelength 
for the fluctuations of the potential.
Although $\lambda$ is not expected to depend on the magnetic field, the scale $Q^{-1}$ and the microscopic mobility $\mu_{xx}$ 
depend the Larmor radius since it gives the characteristic length-scale where the adiabatic theory starts to apply.
In the following we do not try to derive quantitative predictions for these parameters, 
instead we focus on the general properties of Eq.~(\ref{Eq1}) and show that it produces 
a non-linear transport behavior very similar to ZDRS.

Drift diffusion equations in random media has been studied in the past in several contexts 
including fluid dynamics and reaction kinetics in cells \cite{phythian,drummond,king,dean,loverdo}. 
The presence of a random potential $U(\mathbf{r})$ is expected to change the microscopic diffusion rate  
and mobility to effective large scale values $D^*$ and $\mu_{xx}^*$. 
The originality of Eq.~(\ref{Eq1}) comes from the non diagonal mobility tensor ${\hat \mu}$
(in high mobility samples $\mu_{xy} \gg \mu_{xx}$ at even moderate magnetic 
fields $B \simeq 0.1\;{\rm Tesla}$)
and the presence of an applied electric field $\mathbf{E}$ which can distort the disorder potential $U(\mathbf{r})$. As a result we look not only for the effective mobility $\mu_{xx}^*$ but 
for the full dependence of the longitudinal drift velocity $v_x$ on the 
the applied electric field $E_x$.  ZDRS corresponds
to a regime where the drift velocity $v_x$ does not depend on the bias $E_x$. 
Indeed in a Hall bar the current density perpendicular to the channel vanishes
far away from the current injecting electrodes, which leads to an expression for the 
longitudinal field $E_{\parallel} = v_x(E_{\perp})/\mu_{xy}$ as a function of the Hall field 
$E_{\perp} = \frac{I}{R_{H} W_{c}}$ where $I$ the injected current, $R_H$ is the Hall resistance 
and $W_{c}$ is the channel width (simulations on the basis of Eq.~(\ref{Eq1}) and experiments indicate that
the Hall mobility $\mu_{xy}$ and Hall resistance $R_H$ remain almost unchanged by the applied electric field).

\begin{figure}
\centerline{\includegraphics[clip=true,width=8cm]{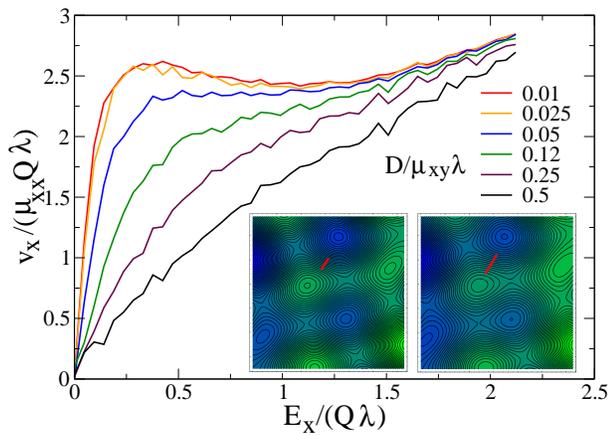}}
\caption{Dependence of the drift velocity $v_x$ as a function of the applied field $E_x$ for several values 
of the microscopic diffusion coefficient $D$ (in dimensionless units). The dynamics of the probe trajectories was 
followed for time $10^6 \mu_{xy} \lambda Q^2$ and $N_q = 100$ disorder Harmonics were used 
in the realization of the disorder potential. The inset show the equi-potential curves for a potential with $N_q = 3$
under electric fields $E_x/(Q \lambda) = 0.25,\; 0.5$ (left/right). The thick red segments indicate the thickness 
of the bundle of delocalized equi-potentials (approximately proportional to $E_x$).
}
\label{SemiFig1}
\end{figure}

The average drift velocity $v_x$ for the model Eq.~(\ref{Eq1}) can be determined 
by integrating the Langevin equations of motion $\mathbf{v} = {\hat \mu} \left( \mathbf{E} - \mathbf{{\rm grad}}\;U \right) + \mathbf{\xi}$ for several probe particles. In this equation $\mathbf{v}$ is the particle velocity and $\mathbf{\xi}$ is a noise term 
whose amplitude is chosen to create the microscopic diffusion rate $D$. In order to simulate the random 
potential, we have taken $U(x,y) = N_k^{-1/2} \sum_{\mathbf{k}} A_{\mathbf{k}} \cos(\mathbf{k} \mathbf{r} + \phi_\mathbf{k})$ 
where the amplitudes $A_\mathbf{k}$ and wavevectors $\mathbf{k}$ were taken from a Gaussian distribution 
with standard deviations $\lambda \sqrt{2}$ and $Q$ respectively (the phases $\phi_\mathbf{k}$ are uniformly distributed 
in $(0, 2\pi)$) . 
The simulations were performed using a Runge-Kutta integration method and typically $N_k = 100$ harmonics 
were used for the realization of the disorder potential. On Fig.~1 we represent the dependence of $v_x$ on the applied 
field, for several values of the microscopic diffusion rate $D$. In absence of the disorder potential 
the drift velocity would be given by $v_{xx} = \mu_{xx} E_x$, this relation is fairly accurate for high values 
of $D / (\mu_{xy} \lambda)$. However when the diffusion rate $D$ is lowered, the dependence $v_x(E_x)$ 
becomes highly nonlinear and starts to exhibit a plateau where $v_x$ depends very weakly on
the applied field $E_x$. For some values of $D$ our simulations suggest a weak decrease of $v_{xx}(E_x)$ (around 5\%)
in this region however the obtained behavior is still remarkably flat. We also noticed 
than the small drop in drift velocity became more pronounced for a small number of Harmonics, 
thus the plateau may become completely flat in the limit $N_k \rightarrow \infty$. 

The origin for this plateau can be understood by representing the equipotentials in presence of 
a moderate applied electric field $E_x$ in the $x$ direction. The potential landscape is formed by closed equipotential 
islands which are separated by narrow streams of delocalized equipotentials which flow perpendicularly (on average) 
to the applied field. The thickness $W$ of these streams grows linearly with the applied field $E_x$ and can be 
estimated as $W \sim E_x Q^{-2} \lambda^{-1}$ \cite{trugman}. Thus the time $T_s \sim W / (\mu_{xx} E_x) \sim Q^{-2} \lambda^{-1} / \mu_{xx}$ 
to drift across the streams is independent of the applied field. After crossing such a stream a carrier is transferred to another island.
In the process it has drifted in the direction of the applied field by a quantity $Q^{-1}$,
leading to a saturation value for the drift velocity: $v_{sat} \sim Q^{-1}/T_s \sim \mu_{xx} \lambda Q^{-1}$. 
This reasoning is plausible as long as the equipotentials are weakly distorted by the applied field
$E_x \ll \lambda Q$; for higher values of $E_x$ we expect to recover $v_x = \mu_{xx} E_x$ as in the absence 
of a disorder potential. Although the above argument yields the correct scaling (as compared to Fig.~1), 
the derivation is arguable in many ways. Thus to pursue analytical investigations we have employed more standard 
field theoretical and renormalization group techniques, following closely the presentation from Ref.~\cite{dean}.

The drift velocity can be obtained from the asymptotic properties of the disorder averaged Green function for Eq.~(\ref{Eq1})
in the limit $\mathbf{k} \rightarrow 0$, indeed $G(\mathbf{k})^{-1} = i (k_x v_x + k_y v_y) + O(k^2)$. 
In the Green-function representation, the perturbation theory series in the power of disorder potential $\lambda$ can be conveniently 
represented through Feyman diagrams. The resulting diagrammatic expansion rules are as follow:
full lines represent the bare Green function in absence of the disorder potential 
\begin{align}
G_0(\mathbf{k}) = \frac{1}{i (\mathbf{k},{\hat \mu} \mathbf{E}) - D k^2} ;
\end{align}
dotted lines carry a factor $\Delta(\mathbf{q})$, and the interaction vertex is $(\mathbf{k}, {\hat \mu} \mathbf{q})$ 
where $k$ and $q$ are the incoming electron/''photon'' wavenumbers respectively. The wavenumbers are conserved 
at each vertex and wavenumber is integrated around closed loops with a factor $d^2 q/(2 \pi)^2$. 

These diagrammatic rules lead us to the following one-loop expression for the free energy (the corresponding diagram is shown on Fig.~2)
\begin{align}
\Sigma(\mathbf{k}) = - \lambda^2 \int \frac{d^2 q}{(2\pi)^2} \Delta(q) (\mathbf{k}, {\hat \mu} \mathbf{q}) (\mathbf{k+q}, {\hat \mu} \mathbf{q}) G_0(\mathbf{k+q}) 
\label{EqSigma}
\end{align}
we remind that the free energy $\Sigma(\mathbf{k})$ is connected to the Green function through the 
Dyson equation $G^{-1}(\mathbf{k}) = G_0^{-1}(\mathbf{k}) - \Sigma(\mathbf{k})$. The correction to the velocity $v_{x}$ is obtained 
by keeping only the linear terms in $k_x$  from Eq.~(\ref{EqSigma}) and neglecting all second order terms in the quantity $\mu_{xx}/\mu_{xy} \ll 1$.
Under this approximation the integral over momentum $\mathbf{q}$ can be computed analytically and leads to:
\begin{align}
v_{x} &= \mu_{xx} E_x + \lambda^2 \frac{\mu_{xx} Q^2}{E_x} R(\frac{\mu_{xy} E_x}{D Q}) 
\label{Eq:Lin}
\end{align}
where the function $R(x)$ is defined as: 
\begin{align*}
R(x) &= 2 - \frac{x^2}{4} e^{x^2/4} \left[ x^2 K_0( x^2/4 ) + (2 - x^2) K_1(x^2/4) \right] 
\end{align*}
with $K_0$ and $K_1$ designating the modified Bessel functions. Although this expression compares favorably 
with the numerical simulations in the limit $\mu_{xy} \lambda / D \rightarrow 0$, 
it introduces a pronounced negative differential resistance state at strong $\lambda$
which is not present in the numerical simulations (see Fig.~2). 

The failure of one loop perturbation theory suggests that higher order nonlinear terms in the disorder 
potential amplitude must also be taken into account.
A possible approach is to use the self-consistent Born approximation where the bare Green function $G_0(\mathbf{k})$ in 
Eq.~(\ref{EqSigma}) is replaced by the full propagator $G(\mathbf{k})$. 
In order to solve the self-consistent equation, we have searched for a solution of the 
form $G(\mathbf{k})^{-1} = i (v_x k_x + v_y k_y) - D^*_x k_x^2 - D^*_y k_y^2$ 
and kept only first and second order terms in wavenumber that stem from Eq.~(\ref{EqSigma}). 
Then we determined $D^*_x$ and $D^*_y$ numerically using the approximation $v_y \simeq -\mu_{xy} E_x$ and $v_x \ll v_y$ 
and used these value to find $v_x$.  Under this approximation, we find
effective diffusion rates $D^*_x$ and $D^*_y$ much stronger than the microscopic value $D$.
This is a consequence of the enhancement of the diffusion rate on large length-scales 
due to the presence of the disorder potential. However for small $D$ the underlying microscopic dynamics 
is almost ballistic at small length-scales, and this fact is missed when the same effective 
diffusion constants $D^*_x$ and $D^*_y$ is used for all length-scales. As a result the self-consistent 
Born approximation leads to a smooth weakly non-linear increase similar to the linear response theory for high $D$ (see Fig.~2)
and does not reproduce the plateau in $v_x(E_x)$. 

\begin{figure}
\centerline{\includegraphics[clip=true,width=8cm]{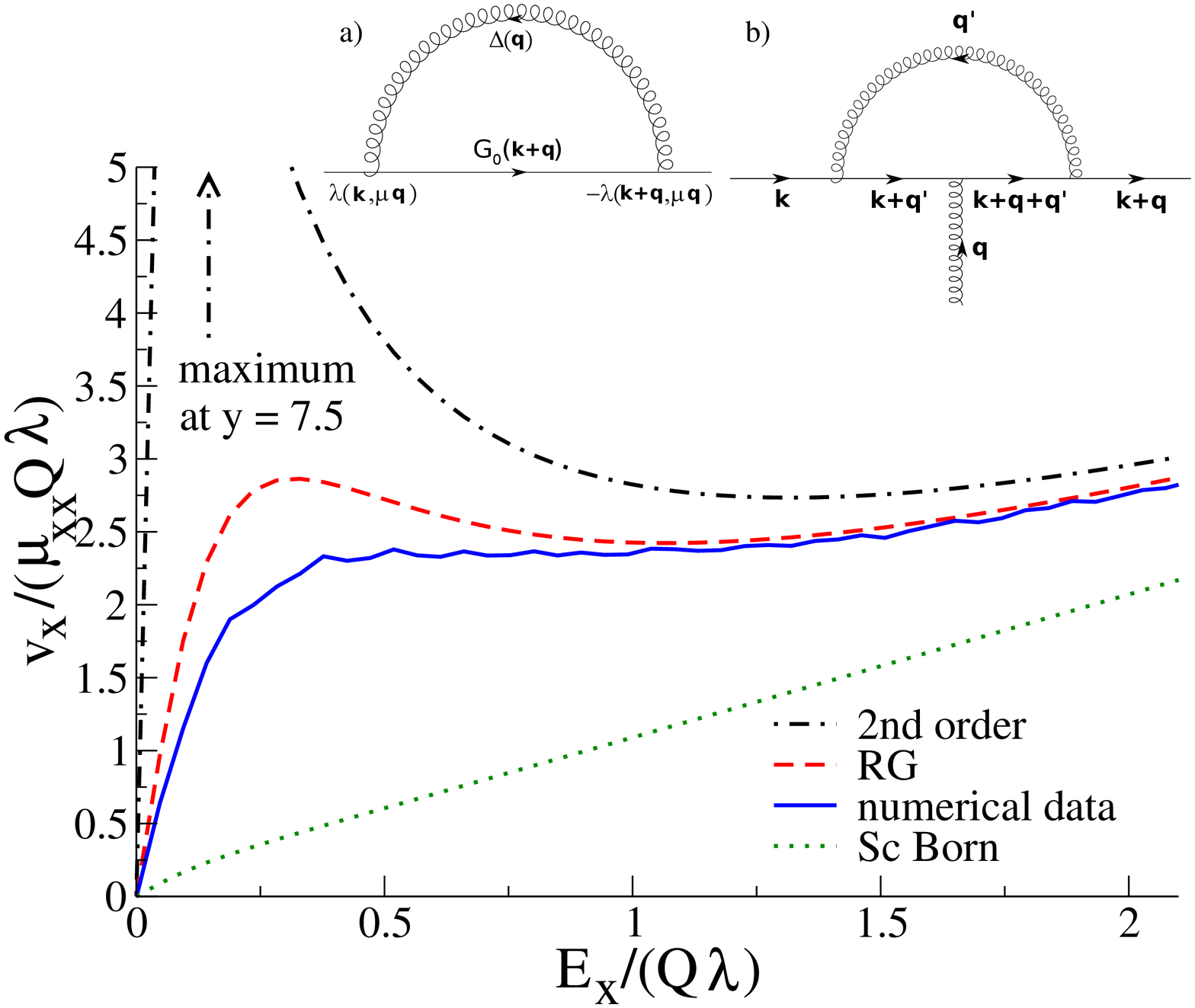}}
\caption{Top figures: a) diagram for the one loop expression of the self energy leading to Eq.~(\ref{EqSigma}) and b) represents the diagram for the vertex correction. The bottom curves show a comparison between the numerical calculation of $v_x(E_x)$ 
and the different analytical approximations 
described in the text (one loop perturbation theory, self consistent Born approximation and renormalization group theory) 
for the case of $D/(\mu_{xy} \lambda) = 0.05$ in Fig.~1.
}
\label{SemiFig2}
\end{figure}

The failure of the self-consistent Born approximation, 
suggests to use a renormalization group (RG) method where $D$ and $\mu_{xx}$ are allowed 
to vary as a function of the length scale $\Lambda^{-1}$. In the absence 
of magnetic and electric fields, \cite{dean} showed that RG
gives an accurate effective diffusion rate. Thus we adapt the 
calculations developed therein to the case of magnetotransport 
in the regime $\mu_{xy} \gg \mu_{xx}$ and in the limit of vanishing electric fields $E_x$. 
In this case the functional form of the Green function at a length scale $\Lambda^{-1}$ 
is simply $G_{\Lambda}(\mathbf{k})^{-1} = - D(\Lambda) k^2$. As in perturbation theory, 
the one-loop diagram allows to compute the change in self-energy $\Sigma_{\Lambda}$ 
when the momentum scale is reduced from $\Lambda$ to $\Lambda-\delta \Lambda$:
\begin{align}
\Sigma_{\Lambda-\delta \Lambda}(\mathbf{k}) - \Sigma_{\Lambda}(\mathbf{k}) =  
\frac{\lambda^2 \mu_{xy}^2}{D(\Lambda)} \mathbf{k}^2 \frac{\Delta(\Lambda)}{4 \pi} \Lambda d\Lambda
\label{dSigma}
\end{align}
The Dyson equation gives a relation between the change in the self-energy and the change 
in the diffusion constant $-d\Sigma_\Lambda/d\Lambda = k^2 d D(\Lambda)/d \Lambda$.
We are thus lead to a first order differential equation on $D(\Lambda)$, 
which after integration gives:
\begin{align}
D(\Lambda) = \sqrt{D^2 + \lambda^2 \mu_{xy}^2 \exp\left(-\frac{\Lambda^2}{2 Q^2} \right)}\;,\;D^*=D(0)
\label{Eq:Deff}
\end{align}
This expression for $D^*$ is in good agreement with 
the numerical simulations in a wide range of parameters. 
The regime $D \rightarrow 0$ deserves special attention; in this limit 
electrons are likely to become trapped around minima of the disorder potential,
which results in vanishing transport coefficients $D^*, \mu_{xx}^*, \mu_{xy}^*$.
Since the RG calculation did not take this possibility into account,
its domain of validity is restricted by $D \gg \lambda \mu_{xx}$
where trapping is not expected.

The effective mobility $\mu_{xx}^*$ 
can be deduced by including the vertex correction represented on 
Fig.~2 in the RG procedure. Indeed in the linear response regime,
$\mu_{xx}$ appears in the diagrams only through the interaction vertex.
The flow equation for $\mu_{xx}(\Lambda)$ is then $d\mu_{xx}(\Lambda) = \mu_{xx}(\Lambda) D(\Lambda)^{-1} d D(\Lambda)$, 
which leads $\mu_{xx}^* = \mu_{xx}(0) = D^* \mu_{xx}/D$. This result can also be
understood intuitively from the Einstein-relation, 
indeed the ratio $D(\Lambda)/\mu_{xx}(\Lambda)$ determines an effective carrier temperature which is 
not expected to depend on the length scale $\Lambda^{-1}$. This expression for the mobility combined with Eq.~(\ref{Eq:Deff}) 
leads to an expression for the mobility which compares favorably with our numerical simulations in the regime 
of weak electric fields. 

In order to generalize the RG calculation to the case of finite electric fields 
several obstacles arise. The drift velocity is determined by the asymptotic properties of the Green function 
and not only by the renormalized interaction vertex; moreover the bare propagator becomes anisotropic which 
makes separation of momentum space in isotropic shells $\Lambda - \delta \Lambda < r < \Lambda$ inappropriate. 
In front of these difficulties we have adopted the following simplifying anzats, we 
assumed that the renormalization flow equation for the drift velocity $v_{x}(\Lambda)$ was of the form 
\begin{align}
\frac{1}{v_{x}(\Lambda)} \frac{d v_{x}(\Lambda)}{d \Lambda} = \frac{1}{D(\Lambda)} \frac{D(\Lambda)}{d \Lambda} S_v\left(\frac{\mu_{xy} E_x}{D(\Lambda) \Lambda} \right)
\label{vx:renorm}
\end{align}
The proposed form is a generalization of the flow equation for the mobility $\mu_{xx}(\Lambda)$.
It incorporates the dependence on the electric field through the scaling function $S_v$ 
which depends on the dimensionless parameter $\mu_{xy} E_x/(D(\Lambda) \Lambda)$.
The choice of this quantity is governed by dimensional analysis and by similarity to Eq.~(\ref{Eq:Lin}).
The connection to Eq.~(\ref{Eq:Lin}) can be exploited even further, by requiring 
that Eq.~(\ref{vx:renorm}) reduces to Eq.~(\ref{Eq:Lin}) in the limit of weak disorder $\lambda \rightarrow 0$.
This allows to find the functional form of $S_v(x)$:
\begin{align}
S_v(x) = \frac{2}{x^2}\left(1 - \frac{1}{\sqrt{x^2+1}}\right)
\label{vx:Sv}
\end{align}
The drift velocity $v_x$ can now be obtained through numerical integration of Eq.~(\ref{vx:renorm})
using the initial condition $v_{x}(\Lambda = \infty) = \mu_{xx} E_x$ and Eqs.~(\ref{Eq:Deff}),(\ref{vx:Sv}).
The obtained values for the drift velocity are in a good agreement with numerical 
simulations (see Fig.~2). They reproduce with a good accuracy the behavior in the limit of low and high electric fields,
however the above approximation also produces a clear negative differential resistance which is 
not present in the simulations. 

Several factors may explain the observed discrepancy.
The change of the diffusion rate by the applied electric field is not captured in Eq.~(\ref{Eq:Deff}).
Under bias the diffusion also becomes anisotropic which requires to introduce two diffusion coefficients 
$D_x(\Lambda)$ and $D_y(\Lambda)$ as in the self-consistent Born approximation.
Moreover the change in $v_x(\Lambda)$ deduced from the one loop expression of the self-energy and 
the change in $\mu_{xx}(\Lambda)$ from the vertex-correction must be treated separately. 
If the isotropic RG procedure is still valid in presence of anisotropy, it may be possible to derive coupled equations 
on the flow of the four quantities: $D_x(\Lambda), D_y(\Lambda), \mu_{xx}(\Lambda)$ and $v_x(\Lambda)$,
however the results obtained from the anzats Eq.~(\ref{vx:renorm}) already show a sufficient agreement with the simulations 
to give confidence in our numerical data and in the truthfulness of the observed the plateau in $v_x(E_x)$.
More advanced RG calculations will not necessarily bring a physical understanding of the plateau origin,
and other analytical methods could be more successful for this purpose \cite{schindler}. 
Thus we did not try to push the accuracy of the RG calculations further. 

So far we did not discuss the role played by the formation of Landau levels under magnetic field.
The first experimental reports \cite{bykov2007} indicated that ZDRS appeared 
only at magnetic fields corresponding to the maximum of a Shubnikov-de Haas (SdH) oscillation,
while the developed theory seem to suggest that they could appear at any magnetic field.
We interpret this apparent contradiction with the experiments as follow: the experiments were conducted 
in a  magnetic field regime where the longitudinal resistance vanished at the minimum of the SdH oscillations.
It is known that in this regime the drop of the Hall voltage along the channel maybe strongly nonuniform 
outside of the SdH maxima \cite{ahlswede,guven} which could lead to an effective smoothing of the ZDRS. 
We note that the recent observation of ZDRS at high filling factors \cite{hatke} before the onset of SdH oscillations 
seems to support this interpretation.

To summarize we have shown through numerical simulations that model Eq.~(\ref{Eq1}) leads to a non 
linear transport behavior very similar to a zero-differential-resistance state without invoking 
an instability argument. Even if our simulations are 
consistent with a weak residual negative-differential resistance, this trend is strongly exaggerated in some 
of the analytical approximations which we employ, specially if only 
the first order disorder contribution to the self energy is kept. 
Approximate renormalization group calculations were the most successful to reach an agreement between
simulation and analytical theory, however even in this case the plateau was not reproduced correctly.
For a quantitative comparison with experiments the parameters of the model must be 
estimated from microscopic transport theory which will be the focus of further attention. 
Moreover edge effects are probably also important, since it is likely that 
edge-trajectories are stabilized against the fluctuations of the disorder potential by a static electric field perpendicular to the edge.
To conclude it is interesting that the proposed simple model already yields a rich non-linear physics 
and can potentially rise an interesting challenge for methods of field theory and non-linear science. 

We acknowledge M. Schindler and D.L. Shepelyansky for fruitful discussions,
K. Kono for his stimulating interest in this work and RIKEN for hospitality during the writing of the manuscript.


\begin{thebibliography}{99}


\bibitem{zudov2001a} M.~A. Zudov,R.~R. Du, J.~A. Simmons, J.~L. Reno, Phys. Rev. B {\bf 64}, 201311(R) (2001)

\bibitem{ye2001}
P.~D. Ye, L.~W. Engel, D.~C. Tsui, J.~A. Simmons, J.~R. Wendt,
G.~A. Vawter and J.~L. Reno, Appl. Phys. Lett. \textbf{79}, 2193 (2001).

\bibitem{yang2002} C. L. Yang, J. Zhang, R. R. Du, J.A. Simmons and J.L Reno {\bf 89}, 076801 (2002) 

\bibitem{zudov2001b}
M.~A. Zudov, I.~V. Ponomarev, A.~L. Efros,
R.~R. Du, J.~A. Simmons and J.~L. Reno,
Phys. Rev. Lett.  \textbf{86}, 3614 (2001).


\bibitem{dorozhkin2005}
S.~I. Dorozhkin,
J.~H. Smet,
V.~Umansky and
K.~ von Klitzing, Phys. Rev. B \textbf{71}, 201306(R) (2005).

\bibitem{studenikin2005}
S.~A. Studenikin, M.~Potemski, A.~Sachrajda,
M.~Hilke, L.~N. Pfeiffer, and K.~W. West,
Phys. Rev. B (\textbf{71}), 245313 (2005).
  
\bibitem{zudov2006} M.A. Zudov, R.R. Du, L.N. Pfeiffer and K.W. West, 
         Phys. Rev. B {\bf 73}, 041303(R) (2006).

\bibitem{mani2010} R. G. Mani, C. Gerl, S. Schmult, W. Wegscheider and V. Umansky, Phys. Rev. B {\bf 81} 125320 (2010) 



\bibitem{ryzhii} V.I. Ryzhii, Sov. Phys. Solid State {\bf 11}, 2078 (1970).

\bibitem{girvin} A.C. Durst, S. Sachdev, N. Read, and S.M. Girvin, Phys.
        Rev. Lett. {\bf 91}, 086803 (2003).

\bibitem{polyakov} I.A. Dmitriev, A.D. Mirlin, and D.G. Polyakov, 
        Phys. Rev. Lett. 
        {\bf 91}, 226802 (2003).

\bibitem{vavilov} M.G. Vavilov and I.L. Aleiner, 
        Phys. Rev. B {\bf 69}, 035303 (2004).

\bibitem{platero} J. I\~narrea and G. Platero, 
        Phys. Rev. Lett. {\bf 94}, 016806 (2005).

\bibitem{platero2} J. I\~narrea and G. Platero, 
        Physica E {\bf 40}, 1902 (2007).

\bibitem{mirlin} I.A. Dmitriev, M.G. Vavilov, I.L. Aleiner, 
        A.D. Mirlin, and D.G. Polyakov, 
        Phys. Rev. B {\bf 71}, 115316 (2005).

\bibitem{monarkha} Y.P. Monarkha, Jour. Low. Temp. Phys. {\bf 37}, 108 (2010)


\bibitem{mani2002}  R.G.~Mani, J.H.~Smet, K. von Klitzing,
         V.~Narayanamurti, W.B.~Johnson, and V.~Umansky,
         Nature {\bf 420}, 646 (2002).

\bibitem{zudov2003}  M.A.~Zudov, R.R.~Du, L.N.~Pfeiffer, and K.W.~West,
        Phys. Rev. Lett. {\bf 90},  046807 (2003).

\bibitem{mani2004} R.G. Mani, V. Narayanamurti, K. von Klitzing, J.H. Smet, 
        W.B. Johnson and V. Umansky Phys. Rev. B {\bf 69}, 161306(R) (2004);
        {\bf ibid.} {\bf 70}, 155310 (2004).

\bibitem{smet} J.H. Smet, B. Gorshunov, C. Jiang, L. Pfeiffer, K. West, V. Umansky, M. Dressel,
R. Meisels, F. Kuchar and K. von Klitzing, PRL {\bf 95}, 116804 (2005)

\bibitem{bykov} A.A. Bykov, A.K. Bakarov, D.R. Islamov and A.I. Toropov, 
        JETP Lett. {\bf 84}, 391 (2006).

\bibitem{bykov2009} A.A. Bykov, JETP Lett. {\bf 89}, 575 (2009)

\bibitem{Wiedmann} S. Wiedmann, G. M. Gusev, O. E. Raichev, A. K. Bakarov, and J. C. Portal 
Phys. Rev. Lett. {\bf 105}, 026804 (2010) 



\bibitem{bykov2007} A.A. Bykov, J-q Zhang, S. Vitkalov, A.K. Kalagin and A.K. Bakarov,
Phys. Rev. Lett. {\bf 99}, 116801 (2007)

\bibitem{zhang2007} J.-q. Zhang, S. Vitkalov, A. A. Bykov, A. K. Kalagin, and A. K.
Bakarov, Phys. Rev. B {\bf 75}, 081305R (2007)

\bibitem{kunold2009} A. Kunold and M. Torres, Phys. Rev. B. {\bf 80}, 205314 (2009)

\bibitem{zudov2010} A. T. Hatke, H.-S. Chiang, M. A. Zudov, L. N. Pfeiffer, and K. W. West 
Phys. Rev. B {\bf 82}, 041304 (2010)

\bibitem{gusev2011} G. M. Gusev, S. Wiedmann, O. E. Raichev, A. K. Bakarov, and J. C. Portal 
Phys. Rev. B {\bf 83}, 041306 (2011) 

\bibitem{bykov2011} A.A. Bykov, E.G. Mozulev and S.A. Vitkalov, JETP Lett. {\bf 92} 475 (2010) 



\bibitem{denis2010} D. Konstantionv and K. Kono, PRL {\bf 105}, 226801 (2010) 

\bibitem{denis2011} D. Konstantionv, A.D. Chepelianskii and K. Kono, arXiv:1101.5667

\bibitem{toulouse} A.D. Chepelianskii and D.L. Shepelyansky, Phys. Rev. B. {\bf 80}, 241308(R), 2009

\bibitem{mikhailov} S. A. Mikhailov, Phys. Rev. B {\bf 83}, 155303 (2011)

\bibitem{toulousecam} A.D. Chepelianskii, J. Laidet, I. Farrer, H.E. Beere, D.A. Ritchie, H. Bouchiat, arXiv:1102.2314


\bibitem{chirikov} B.V. Chirikov, Phys. Rep. {\bf 52}, 263 (1979).

\bibitem{phythian} R. Phythian and W.~D.~Curtis, J. Fluid Mech. {\bf 89}, 241 (1978)

\bibitem{drummond} J.T. Drummond, S. Duane and R.R. Horgan, J. Fluid Mech. {\bf 138}, 75 (1984)

\bibitem{king} P.~R.~King, J. Phys. A: Math Gen {\bf 20}, 4661 (1987)

\bibitem{dean} D.~S. Dean, I.~T. Drummond and R.~R. Horgan, J. Phys. A:Math. Gen. {\bf 27}, 5135 (1994)

\bibitem{loverdo} C. Loverdo, O. B\'enichou, M. Moreau and R. Voituriez, Nature Phys. {\bf 4}, 135 (2008)

\bibitem{trugman} S.A. Trugman, Phys. Rev. B {\bf 27}, 7539 (1983)

\bibitem{schindler} M. Schindler in preparation (2011)


\bibitem{ahlswede} E. Ahlswede, P. Weitz, J. Weis, K. v. Klitzing and K. Eberl, Physica B {\bf 298}, 562 (2001)

\bibitem{guven} K. G\"uven and R.R. Gerhards, Phys. Rev. B {\bf 67}, 115327 (2003)

\bibitem{hatke} A.T. Hatke, M.A. Zudov, L.N. Pfeiffer and K.W. West, Phys. Rev. B {\bf 83}, 081301 (2011)
\end{thebibliography}
\end{document}